\documentclass[useAMS,usenatbib]{mn2e}

\usepackage{graphicx}
\usepackage{multirow} 
\usepackage{amssymb,amsmath}
\title[Variability from Circumbinary Discs]{Infrared Variability from Circumbinary Disc Temperature Modulations}
\author[Bodman \& Quillen]{Eva H.L. Bodman, \& Alice Quillen \\
Department of Physics and Astronomy, University of Rochester, Rochester, NY 14627, USA; \\
}
\date{} 

\begin{document}

\maketitle

\begin{abstract}
The temperature of a circumbinary disc edge should undulate due to variations in illumination as a function of binary orbital phase. We explore circumbinary disc temperature variations as a source of broad-band infrared light curve variability. Approximating the wall of a circumbinary disc edge as a wide optically thick cylinder with surface temperature dependent on its illumination, we find that a binary comprised of  1 $M_\odot$ and 0.5 $M_\odot$ pre-main sequence stars in a $\sim$15.5 day period, would exhibit the largest amplitude variations of $\sim$9\% at 3.77 and 4.68 $\mu$m as seen by a distant  observer. The amplitude of variations and shape of the light curve is sensitive to the luminosity and mass ratios of the stars in the binary, the radius of the circumbinary disc clearing, the binary separation, and the orbital inclination. The light curve variations are smooth and very red with a non-sinusoidal shape for most of the parameter space explored. Possible morphologies include a single peak with a flat region, two peaks of different heights or a single dip. 
\end{abstract}

\section{Introduction}

Variability is ubiquitous among young stellar objects (YSOs) and occurs on a variety of time-scales and over a wide range of wavelengths, from radio through X-rays. Near and mid-infrared monitoring campaigns probe relatively cool material that is dim at shorter wavelengths, and are sensitive to distribution,  temperature, and opacity changes in circumstellar gas and dust. Periodic variations in photometric measurements can be associated with rotational modulation of cool star spots or hot accretion spots,  stellar pulsation, transits, eclipses and intervening circumstellar material \citep[e.g.,][]{FE1996, CHS2001, WHJ2004, vJH2010, MSH2011, WRA2013, CSB2014, PPWG2014}.

Most young stars are born in binary star systems \citep{DK2013}  and in these, additional processes contribute to periodic light curve modulations including eclipses, ellipsoidal tidal deformations of the stars, reflection of light of each star from its companion and at low amplitude, relativistic beaming \citep{ZMA2007,TFM2015}. A binary star system can be embedded in a circumbinary disc \citep[e.g.,][]{Jensen2007,IK2008, N2012,Biller2012, Takakuwa2013,Gillen2014}.  Periodic broad-band photometric variations  could constrain the location or shape of the circumbinary disc edge \citep{N2010,N2012},  infer the presence of a low mass companion \citep{Demidova2013,Maiz2015} or discover a circumbinary or circumsecondary disc \citep{Rattenbury2015}. As studies increase light curve quality, the number of data points, and the wavelength coverage, it may become possible to differentiate between different variability mechanisms using light curve morphology and colour. Circumbinary discs are of particular interest as recently discovered circumbinary planets imply that circumbinary discs host planet formation \citep{MT2015}.

A possible source of IR variability is temperature oscillations in a circumbinary disc due to uneven irradiation from the binary star system (see \citealt{N2010}). The circumbinary disc centre is expected to be at the binary's centre of mass for circular orbits. The disc can contain an inner clearing or cavity caused by the binary \citep{AL1994,PSA2005,PSA2008}. Since neither star in a binary is located at the centre of the disc, the distance between one of the stars and a point in the inner edge or wall of a circumbinary disc depends on the binary's orbital phase. The oscillations in distance cause a modulation in the absorbed stellar flux locally in the disc which leads to a local disc wall temperature dependent on binary orbital phase and azimuthal angle (see Figure \ref{fig:orbit}). \cite{N2010} studied the effects of a binary star on the temperature of the inner wall of an optically thick circumbinary disc and the impact on the spectral energy distribution (SED) for the binary transition disc system CoKu Tau/4. The authors predict variability in the SED with the orbital period, however, because of the long orbital period (about 20 years) for CoKu Tau/4's binary they did not predict the shape of the light curve.
\citet{SK2015} and \citet{C2013} investigated the effects of a binary star on the ice line in a circumbinary disc and focused on wider binaries. In both cases, temperature variations are due to the oscillations in distance from each star to the disc wall.
 
In this paper, we explore how temperature variations in a circumbinary disc would introduce variability in a light curve. Our goal here is to estimate the amplitude, shape and colour of variations in a light curve that are caused by modulation in the temperature of an optically thick wall of a circumbinary disc. In Section \ref{sec:model}, we describe our model and how we compute infrared light curves. In Section \ref{sec:mag}, we discuss the amplitude of the magnitude variations predicted by our model and how the amplitude is effected by the stellar properties of the central star and the geometry of the disc. We describe the sensitivity of the light curves to the stellar luminosity and mass ratios, orbital separation, circumbinary disc wall radius and orientation or viewing inclination. A summary and discussion follows in Section \ref{sec:con}.

 \section{The model} \label{sec:model}

\begin{figure}
\includegraphics[width=3.0in, trim= 0 0 0 0 ]{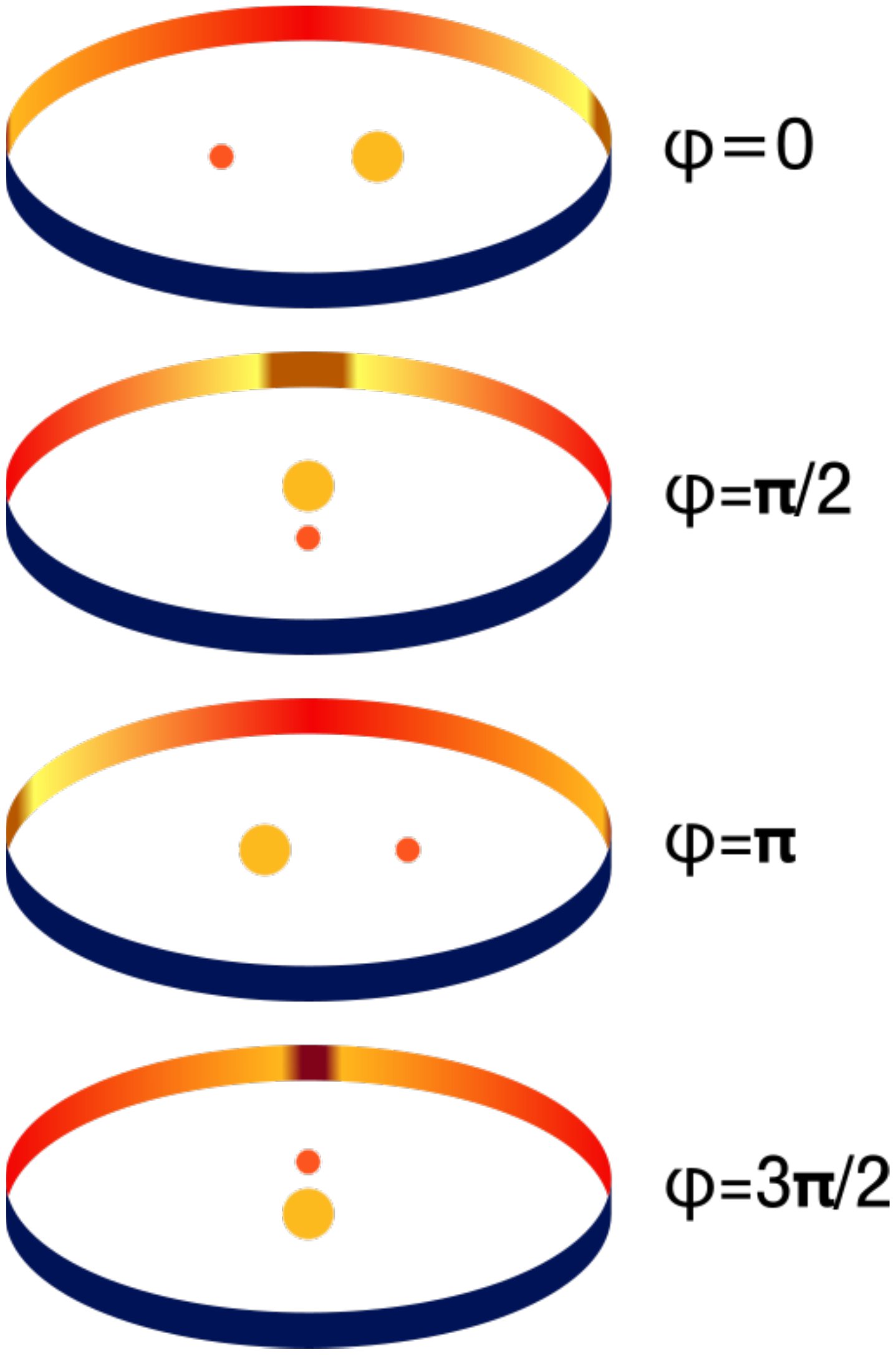}
\caption{Diagram of the temperature variation along a circumbinary disc wall as it varies during the binary orbit. The disc wall is shown as a cylinder with colour showing its local temperature. Yellow is hotter than red which is hotter than brown. Orbital phase is denoted by $\phi$.  The disc wall is assumed to be optically thick so we have neglected radiation from its back side (shown in navy blue).  The disc outside the wall is not drawn. At $\phi=\pm \pi/2$, one star can shadow the other giving a cooler (brown) region on the disc wall. The primary, larger star (shown in yellow) is more luminous and the disc wall is hotter near this star. The observed flux at $\phi =\pi$ is the same as at $\phi=3\pi/2$ but that at $\phi=0$ is not the same as that at $ \phi =\pi$.
}
\label{fig:orbit}
\end{figure} 

\begin{figure}  
\includegraphics[width=3.0in, trim= 0 0 0 0 ]{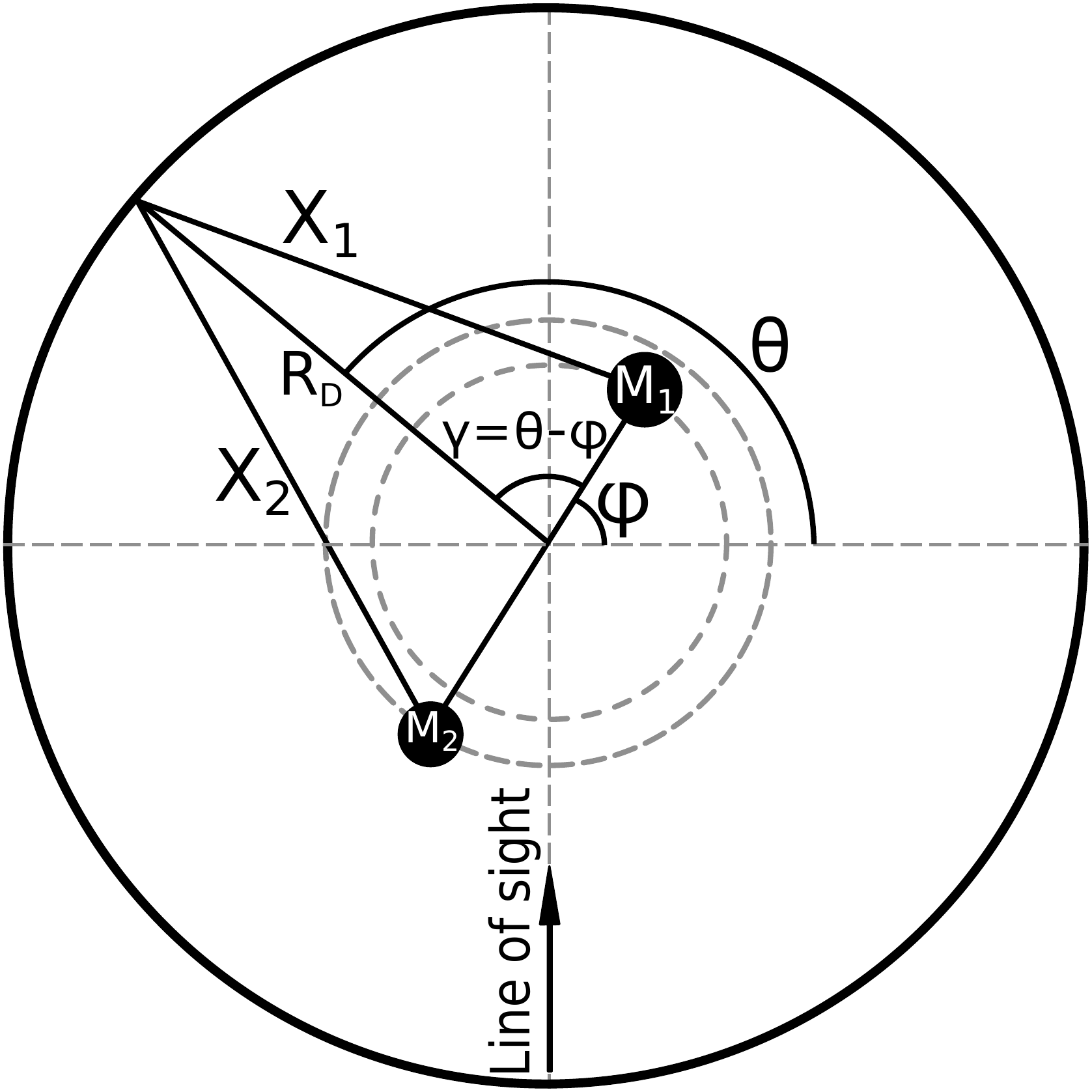}
\caption{Schematic top-down diagram of binary system with circumbinary disc illustrating the geometry of our model. The primary and secondary stars are labelled with mass $M_1$, $M_2$, and have semi-major axes $a_1,\ a_2$ and zero eccentricity. The circumbinary disc wall radius is $R_D$.  The azimuthal angle $\theta$ gives a location in the disc wall with $\theta=0$ normal to the line of sight. The orbital phase is described with $\phi = nt$ with $\phi=0$ when the binary is perpendicular to the line of sight. Here $n$ is the mean motion of the binary. $X_1(\theta,\phi)$ and $X_2(\theta,\phi)$ refer to distances between a location in the disc wall and the primary and secondary stars, respectively. The line of sight is towards the top of the page. We define an angle $\gamma = \theta- \phi$ that helps us compute the distances $X_1,X_2$. 
}
\label{fig:topdown}
\end{figure} 

We first describe the geometry of our model and then we describe how we integrate the flux from the disc wall. Our simple model consists of a  circumbinary disc wall and two stars in circular orbits (see Figure \ref{fig:orbit}). We assume that the disc is not inclined with respect to the orbital plane. The disc wall is a cylinder characterised by its vertical height, $H$, measured from the orbital mid-plane and its radius, $R_D$ measured from the centre of mass of the binary. 
We make no assumptions about the composition of the disc wall other than it is optically thick and is well described as a blackbody with temperature dependent on azimuthal angle, $\theta$ (see Figure \ref{fig:topdown}), neglecting temperature variations as a function of height above the orbital plane. We are neglecting dust features and only using a blackbody spectrum to describe the disc wall emissions. The dust features are expected to change the size of the light curve variations only by a factor of a couple (see Section \ref{sec:coku}).
We do not expect an optically {\it thin} disc to exhibit light curve modulations from temperature variations as the 
total emitted disc flux would be independent of viewing angle.  We assume the stars are approximated with blackbodies and that the binary orbit and circumbinary disc wall are circular. Both stars irradiate the disc but since neither star is at the centre of the disc, the distance between star and different points along the wall varies so the absorbed flux and thus the local temperature varies as a function of azimuthal angle and orbital phase, $\phi =  nt$ with $\phi=0$ with stars oriented on a line perpendicular to the line of sight and $n$ the mean motion of the binary. At $\phi=0$, the primary star is located at $\theta=0$ and the secondary star at $\theta=\pi$ so the primary and secondary eclipses, if observable, would occur when $\phi=\pi/2$ and $3\pi/2$, respectively. As the stars travel along their orbits, the distance between each star and a point on the disc wall varies causing the observed flux from that region to oscillate. The half of the disc wall that contributes to the observed disc flux is the range $\theta = 0$ to $\pi$.  

Each star is characterised by its effective temperature, $T_{\text{eff},i}$, radius, $R_{i}$,  mass, $M_{i}$, luminosity, $L_{\star,i}$,  and semi-major axis,  $a_i$,  with index $i=1,2$ depending on the star. Their SEDs are given by blackbody curves.  The separation between the two stars we denote $a_{\text{sep}}$. We neglect stellar eclipses, ellipsoidal deformation, reflection and relativistic beaming. We assume that the disc wall reaches thermal equilibrium on a time-scale much shorter than the orbital period. Following \cite{N2010}, the thermal time-scale is 
\begin{equation}
t_\text{thermal}={k\Sigma_R\over 2 m_H \sigma T^3}
\end{equation}
where $k$ is the Boltzmann constant, $m_H$ is the mass of Hydrogen, $\sigma$ is the Stefan-Boltzmann constant, $T$ is the temperature of the wall. The surface density of the wall is  $\Sigma_R=\int^{R_D+dR}_{R_D}\rho dR$ and the volumetric density is estimated by $\rho=\Sigma_d/H$. 
Note we are interested in the surface parallel to the disc wall and $\rho$ is estimated with the surface density of the top of the disc. The surface density of a disc at about 0.1 AU is $\Sigma_d\approx$ 500 g cm$^{-2}$ \citep{DCL1998}. Since the wall emission come from a thin radial region of the disc, we estimate the wall surface density by setting $\Delta R= H$ such that $\Sigma_R=\Sigma_d$. The thermal time-scale is then
\begin{equation}
t_\text{thermal}\approx5.4\left(\Sigma_R\over500\, \text{g cm}^{-2}\right)\left(T\over920\,\text{K}\right)^{-3}\text{days}.
\end{equation}
The thermal time-scale is comparable to the orbital time-scale only for very close binaries. 

The visible part of the disc wall is the side facing the observer and the other side is assumed to be completely occulted by absorption of the outer regions of the disc (see Figure \ref{fig:orbit}). We model the disc wall as an optically thick cylinder with vertical walls and we neglect radiative transfer effects. All of the stellar radiation that reaches the wall is absorbed and reemitted at the surface. The disc wall is the warmest part of the disc and it receives a sizeable portion of the stellar irradiation. Consequently emission from the disc wall dominates the spectral energy distribution and we neglect emission from the rest of the disc.
 
Assuming the disc wall has an albedo of zero, the absorbed stellar flux at a given azimuthal angle $\theta$ on the disc wall is 
\begin{equation}
F_\text{Wall}(\theta,\phi) = \left({L_{\star,1}\over 4\pi X_1^2(\theta,\phi)}+{L_{\star,2} \over 4\pi X_2^2(\theta,\phi)}\right)
\label{eq:lum}
\end{equation}
when one star is not blocking light from the other star to the disc wall. The absorbed flux also depends on the orbital phase of the binary as the distances to the stars also depend on this angle. Here $X_i(\theta,\phi)$ is the distance between the $i$-th star and the point on the wall with azimuthal angle $\theta$ in the orbital plane (see Figure \ref{fig:topdown}) and during orbital phase $\phi$. It is convenient to define an  angle 
 \begin{equation}
 \gamma \equiv \theta-\phi \label{eqn:gamma}
 \end{equation}
After choosing the semi-major axes, the distances $X_i$ depend only on the angle $\gamma$ so henceforth we write $X_i(\gamma)$. We use  small angle approximations so that the flux is independent of height and the temperature variation as a function of height can be neglected.  This is valid when $H\ll R_D$. We also assume that the stellar flux  is normal to the surface of the wall and this is valid when radii $R_{\star,i}\ll X_i$.

\begin{figure}
\includegraphics[width=3.0in, trim= 0 0 0 0 ]{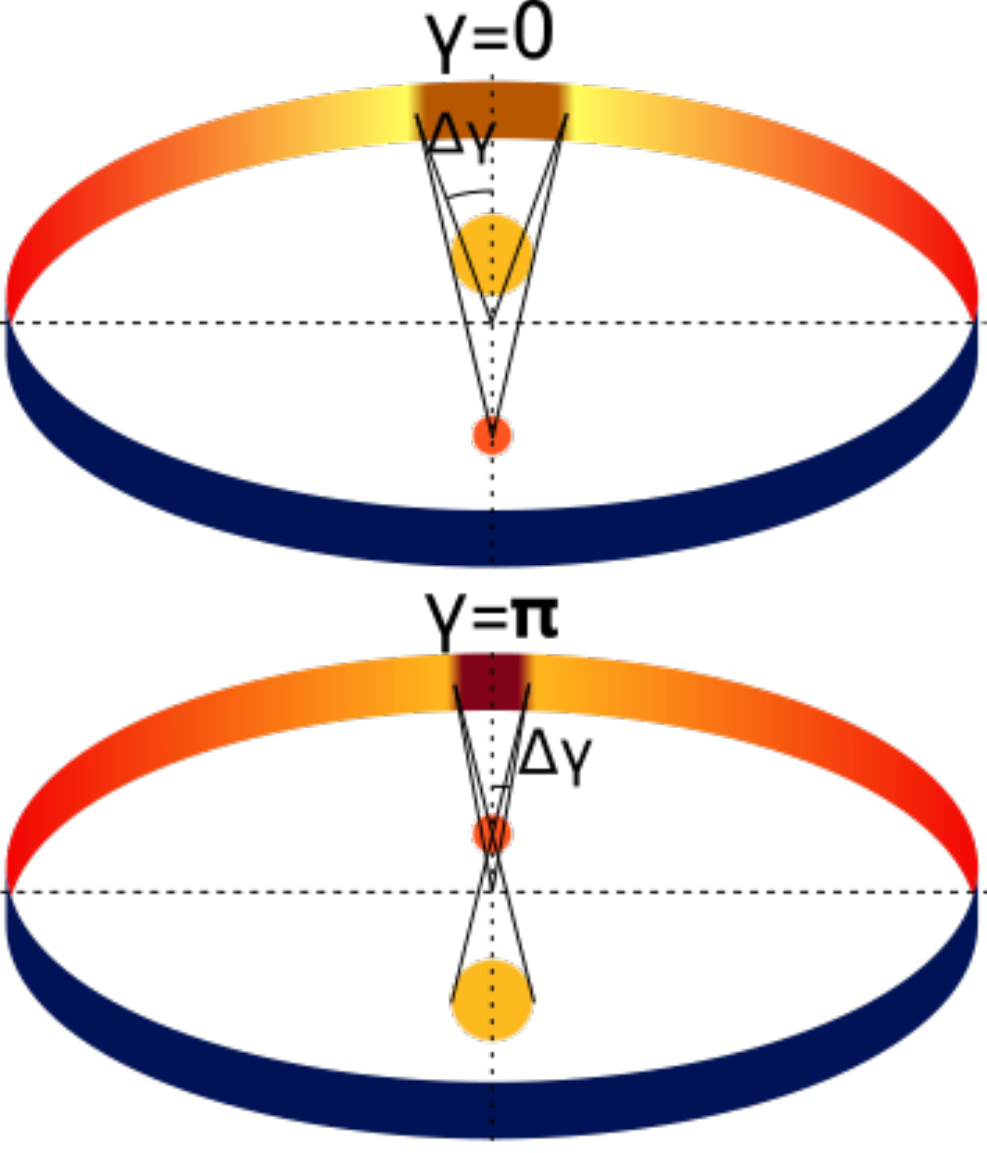}
\caption{Diagram of the shadowed regions from the secondary eclipse on top and primary eclipse on bottom. The colours variance along the disc wall indicates the local temperature in the same manner as in Figure \ref{fig:orbit}. Yellow is hotter than red and navy is the back side where radiation is neglected. The brown colour shows the cooler temperatures from the stellar eclipses. The larger yellow star is the primary and the smaller red star is secondary. The lines mark the geometry used to calculate the width of the shadow which ignores the radius of the smaller star. The top diagram shows the shadow from the secondary eclipse (see equation \ref{eq:ecl1}) and the bottom diagram shows the primary eclipse shadow (see equation \ref{eq:ecl4}). This approximation gives the shadowed region sharp edges in the centre of the smooth transitions shown in the diagram. The disc outside the wall is not shown.
}
\label{fig:shadow}
\end{figure}

When one stars occults the other star, as seen from the disc, the disc temperature can be reduced (see Figure \ref{fig:shadow}). Since the disc surrounds the binary star and is aligned with the binary orbit, there will be two regions along the disc wall where one stars shadows the other, as seen from a particular position on the disc wall. The shadows cause sharp dips in temperature and they are illustrated as brown regions in Figure \ref{fig:shadow}. The primary's shadow occurs at $\gamma=0$ while the secondary's is at $\gamma=\pi$. We first discuss the angular width of the primary shadowing the secondary and then the angular width of the secondary shadowing the primary.

We first consider the situation where the primary is larger than the secondary. In this case, light from the secondary does not reach the disc wall during secondary eclipses. The shadow on the disc has an angular width $\Delta\gamma$ measured from the centre of the shadow where
\begin{equation}
\left|\tan\gamma\right|\leq\tan(\Delta\gamma)= {R_{1}\over R_D} {R_D+a_2\over a_\text{sep}}
\label{eq:ecl1}
\end{equation}
as illustrated by the top drawing in Figure \ref{fig:shadow}. We denote the separation between the stars as $a_\text{sep}$. The shadow where light from the primary is blocked by the secondary has a width of  
\begin{equation}
\left|\tan(\pi-\gamma)\right|\leq\tan(\Delta\gamma)= {R_{1}\over R_D}{R_D-a_2\over a_\text{sep}}.
\label{eq:ecl4}
\end{equation} 
In this region the primary is only partially occulted by the smaller secondary and the shadow is smaller than the primary's shadow (see bottom of Figure \ref{fig:shadow}). 

If the secondary is larger than the primary,\footnote{This does not occur for coevolving PMS stars studied here but may occur for more evolved systems.} then the primary partially occults the secondary where 
\begin{equation}
\left|\tan\gamma\right|\leq\tan(\Delta\gamma) = {R_{2}\over R_D} {R_D-a_1\over a_\text{sep}}
\label{eq:ecl3}
\end{equation}
and the primary is completely shadowed by the secondary with
 \begin{equation}
\left| \tan(\pi-\gamma)\right|\leq\tan(\Delta\gamma) = {R_{2}\over R_D}{R_D+a_1\over a_\text{sep}}.
 \label{eq:ecl2}
 \end{equation}

In Figure \ref{fig:shadow}, the width of the shadow is larger than the height of the disc but for stars with small radii or for high disc walls, that is not true. Here we neglect the height of the shadows, assuming that the shadow covers the full disc height for all star radii and disc heights. Since we are integrating over an area much larger than the shadow, the transition or partially shadowed regions can be neglected. We assume that shadow region begins when the half of one star occults the other giving a sharp transition from no shadow to full shadow. 
 
For angle $\gamma$ satisfying equations \ref{eq:ecl1} and \ref{eq:ecl2}, the disc wall only receives light from the primary or the secondary, respectively.  In these shadows, the absorbed stellar flux is 
$$F_\text{Wall,c}= \frac{L_{\star,\ell}}{ 4\pi X_\ell^2}$$ 
where $\ell$ refers to whichever star is larger. 
For angles satisfying equations \ref{eq:ecl4} or \ref{eq:ecl3}, the disc wall absorbs a flux
 \begin{eqnarray}
 F_\text{Wall,p}&=& {L_s\over 4\pi X_s(\gamma)^2} \nonumber\\
 &&+{L_\ell\over 4\pi X_\ell(\gamma)^2}\left(1-{R_s^2\over R_\ell^2}\left(R_D+a_\ell\over R_D-a_s\right)^2\right)
 \end{eqnarray}
 where $s$ denotes the smaller star. Since the smaller star is much closer to the disk edge than the larger star, the smaller can completely occult the larger star when $R_\ell<R_s\times (R_D+a_\ell)/(R_D-a_s)$. In this case, the flux is the same as $F_\text{Wall,c}$ but with $s$ replacing $\ell$.
 
 Assuming that the absorbed flux is reradiated thermally, the local mid-plane disc wall temperature outside of shadows 
(none of the conditions in equations \ref{eq:ecl1} - \ref{eq:ecl3} is satisfied) is given by
 \begin{eqnarray}
T(\gamma)&=& \left({F_\text{Wall}(\gamma )\over F_\odot}\right)^\frac{1}{4} \nonumber\\
&=& 5778\,\text{K} \left({L_{\star,1}\over X_1^2(\gamma)}+{L_{\star,2} \over X_2^2(\gamma)}\right)^\frac{1}{4}
\label{eq:temp}
 \end{eqnarray}
where $F_\odot=L_\odot/4\pi R_\odot^2$ is the surface flux of the Sun, $L_{\star,i}$ is in solar luminosities and $X_i$ is in solar radii and we have assumed a wall emissivity of 1.  For regions of complete and partial shadow, $F_\text{Wall}$ is replaced with $F_\text{Wall,c}$ and $F_\text{Wall,p}$, respectively. 

The azimuthal temperature profile has two peaks corresponding to the regions along the disc wall closest to the primary at $\gamma =0$ and the secondary at $\gamma=\pi$. The relative sizes of the peaks depend on the primary to secondary luminosity ratio, $L_{\star,1}/L_{\star,2}$. The closer the ratio is to one, the closer to equal height the peaks are with the primary peak higher. For two identical stars the peaks at $\gamma=0,\pi$ would be the same height and temperature oscillations would occur at twice the orbital period along with the light curve modulations. The secondary peak disappears at large luminosity ratios as the primary dominates the stellar irradiation of the disc wall. The primary peak is lower for very large primary to secondary mass ratios, $M_1/M_2$. When the primary is much more massive than the secondary, its orbit is small (from the centre of mass) so that its distance to the disc wall does not vary much. Sharp dips in temperature occur in the centre of each peak from each star  shadowing the other and these  occur at $\gamma=0,\pi$. 
 
To model the observed light curve, the local disc flux is calculated using Planck's function $B_\lambda(T(\theta,\phi),\lambda)$ and the local disc wall temperature.  We compute the observed monochromatic flux from the disc wall as a function of wavelength, binary phase $\phi$, and orbital inclination $i$ by integrating the local surface flux over the disc wall area
 \begin{eqnarray}
 F(\phi,i)_{\lambda.\text{Disc}}  &= & {2 R_D\over d^2} \int_0^\pi B_\lambda(T(\theta,\phi),\lambda) \nonumber \\
 &&\times  H_\text{eff}(\theta,i)\sin i \sin \theta d\theta
 \end{eqnarray} 
where $d$ is the distance to the disc from the observer and inclination $i$ with an edge-on orbit having $i= 90^\circ$. $H_\text{eff}(\theta,i)$  is the effective disc height taking into account occultation of the wall by the nearer side of the disc as seen by the observer. Since the disc wall is optically thick, we assume the no radiation from disc wall reaches the observer in the regions obscured by the opposing side of the disc. The amount the nearer half of the disc obscures the wall depends not only on the inclination but also the azimuthal angle along the wall, $\theta$. The height of the projected hole in the disc is smaller near $\theta=0,\pi$ so more of the inner disc wall will be occulted. The disc is locally occulting itself if
 \begin{equation}
 R_D\cot i\sin\theta < H.
 \end{equation}
When true, $H_\text{eff}=R_D\cot i\sin\theta$ but otherwise the full disc wall height $H$ is used. This self-occultation of the disc 
only gives a large effect at inclinations very close to $90^\circ$ since we assume that the disc radius is much larger than the disc 
scale height. Lastly, we add the flux from each star to the disc flux to calculate the total flux for the light curve,
\begin{equation}
F(\phi,i)_{\lambda,\text{total}}=F(\phi,i)_{\lambda.\text{Disc}} +F_{\lambda,1}+F_{\lambda,2}.
\end{equation}
Since the disc flux increases with disc temperature, there are peaks in the light curve at binary phases that give peaks in the temperature at $\theta=\pi/2$. The light curve peak from the primary occurs at binary phase $\phi=\pi/2$. If the secondary temperature peak is large enough, there is also peak in the  light curve peak at binary phase $\phi =3\pi/2$.

The shadows from primary and secondary typically have a small effect on the light curve. The shadows change the shape of the light curve slightly near $\phi=\pi/2$ and $3\pi/2$. The regions of lower temperature occur at the peaks in the temperature and decrease the observed disc wall flux the most when the temperature peaks are at $\theta=\pi/2$. This decreases the peaks in the light curve and increases dips that occur at $\phi=\pi/2,3\pi/2$. Occultation from the near disc wall does not change the shape but deceases the light curve peaks at very high inclinations.
 
Three additional sources of occultation are neglected: the star occulting the disc wall, the disc occulting the star, and the stellar eclipses (one star eclipsing the other). The disc eclipsing the wall and the wall eclipsing the star occur approximately at inclinations larger than 80$^\circ$ but the exact value depends on the geometry of the system like the disc radius and the binary separation. Stellar eclipses typically occur at higher inclinations than the disc eclipses. The decrease in observed flux from stellar eclipse or disc occulting the star is typically much larger than the variations from temperature oscillations since they are of order of the stellar flux, see Table \ref{tab:compare}.

 \section{Fiducial Model}\label{sec:mag}
 
\begin{table}
\begin{center}
\vbox to 65mm{\vfil
\caption{\large Fiduciary Model Parameters}
\begin{tabular}{@{}ll}
\hline
$M_1$  & 1.0$M_\odot$   \\
$M_2$   & 0.5$M_\odot$       \\
$R_1$ & 1.935 $R_\odot$ \\
$R_2$  & 1.461 $R_\odot$   \\
$T_{\text{eff},1}$ & 4417K \\
$T_{\text{eff},2}$ & 3932K\\
$a_\text{sep}$ &30 $R_\odot$ $\approx$0.14AU \\
$i$ & 75 deg \\
$R_D/a_\text{sep}$ & 1.8 \\
$H/R_D$ & 0.0960 \\
\hline
\end{tabular}
{\\Fiducial model parameters for temperature profile and light curves in figure \ref{fig:lcwave}. Stellar mass, radius and surface temperature are consistent with 2 Myr old pre-main-sequences stellar evolution models by \cite{S2000}. \\
\label{tab:model}}
\vfil}
\end{center}
\end{table}

We begin our study of the light curve with a fiducial model and then we vary each model parameter from the fiducial model values individually. Our model is an approximate representation of a binary star with an optically thick circumbinary disc and such discs can be found around PMS stars. Young PMS stars typically have an optical thick inner region in their proto-planetary discs that extend outward from the star to about 10 AU \citep{WC2011}. We choose a fiducial model consisting of two PMS stars of the same age and a circumbinary disc. We set the age to 2 Myrs which is well within the disc half-life of 5-6 Myrs \citep{B2013}. We use PMS stellar evolution models from \cite{S2000} to estimate the properties of a 1.0 $M_\odot$ primary and 0.5 $M_\odot$ secondary.  We choose a binary separation of 30 $R_\odot\approx$0.14 AU which sets the orbital period to about 15.5 days. The inclination is 75$^\circ$ which is low enough for neither star to eclipse the disc or the disc to eclipse either star.  The values for the stellar and orbital properties used in our fiducial model are listed in table \ref{tab:model}.

The disc radius and height can be constrained with dynamical arguments. For a circumbinary disc, a hole in the centre is created through dynamical interactions with the binary. \cite{AL1994} found the minimum radius for the wall of a circumbinary disc through a hydrodynamical approach that is proportional to the binary separation and increases with eccentricity. For zero eccentricity orbits, the authors found $R_D=1.8a_\text{sep}$ which we use for our fiducial model. With a separation of 30 $R_\odot$, the inner disc radius is 54 $R_\odot\approx$0.25 AU. We estimated the height of the disc wall from hydrostatic equilibrium condition, $h\approx c_s / \Omega$ where $\Omega$ is the angular velocity and $c_s\approx\sqrt{kT/\mu}$ is the sound speed with $\mu$ as the mean particle mass and set $H/h=2$. This value of $H/h$ is consistent with the values used in \cite{DHC2005}.
We use a single height for the disc wall by calculating $h$ with the average temperature and for our fiducial model, $H\approx 5.2\ R_\odot$. With this value for $H$, the width of region in the primary eclipse shadow is about the size the disc wall height and the width for the secondary eclipse is smaller than the scale-height so straight edges is a poor approximation. Since the shadows of the stars on the disc don't strongly affect the light curve, this neglect should not strongly  affect our conclusions. 

For ease of comparing the light curves, the flux for each light curve is converted into a magnitude and then the magnitude of each light curve is scaled so that the magnitude at $\phi=0$ for every light curve is exactly equal to one. Since the magnitude is scaled, the distance to the system, $d$, is arbitrary. 

\begin{figure*}
\includegraphics[width=7.0in, trim= 0 0 0 0 ]{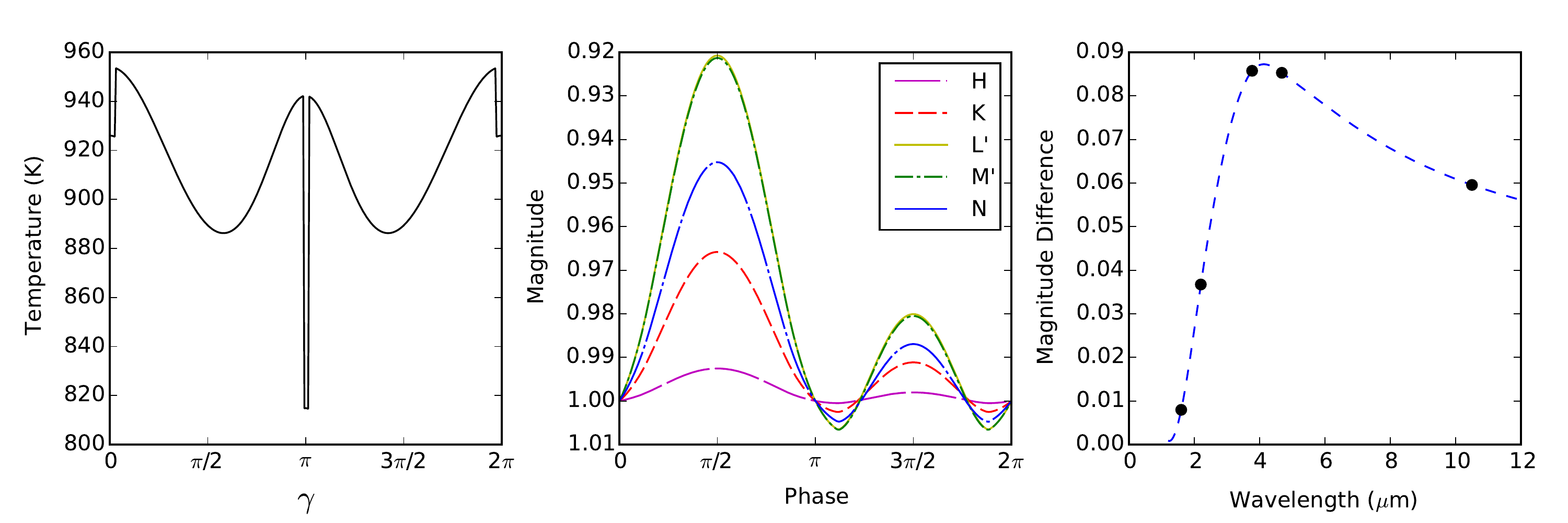}
\caption{Fiducial binary and circumbinary disc model. The left panel shows the azimuthal temperature profile of the inner disc wall. The centre panel gives the light curves at five different wavelengths. Standard IR filter wavelengths were chosen: long dashed magenta for H (1.6 $\mu$m), dashed red for K (2.2 $\mu$m), solid yellow for L' (3.77 $\mu$m), dash-dot green for M' (4.68 $\mu$m), long dash-dot blue for N (10.5 $\mu$m). The right panel is the peak to peak magnitude difference at different wavelengths.  Parameters for the model are listed in Table \ref{tab:model}.
}
\label{fig:lcwave}
\end{figure*}

For the fiducial model, Figure \ref{fig:lcwave} shows the azimuthal temperature profile on the left panel and the light curves at various wavelengths in the middle panel. The light curve has a smooth non-sinusoidal shape. The luminosity ratio of the primary to the secondary is about 3 so the peak in the temperature from the secondary is nearly the same size as the primary peak. The secondary temperature peak is large enough to cause a second peak in the light curve that is about one quarter the size of the primary peak, depending on the wavelength. The light curve reaches a maximum when the primary is closest to the centre of the observable area of the inner disc wall at $\phi=\pi/2$. The light curve reaches a minimum after $\phi=\pi$ and again before $\phi=2\pi$ since there is a smaller peak at $\phi=3\pi/2$ when the secondary is close to the midpoint of the observable side of the disc.  Despite our care in computing shadows in the temperature profile they do not significantly affect the light curves.

The size of the light curve peak depends on the wavelength observed. For the fiducial model the size of the magnitude difference, $\Delta m=m_\text{max}-m_\text{min}$, in M' band (4.68 $\mu$m) is about 0.08 mag. The peak shrinks quickly with decreasing wavelength from 4 $\mu m$ even though the peak wavelength for a 900 K blackbody is $\sim3.2\ \mu$m. At these shorter wavelengths, the light contribution from the stars is much larger so the variability of the disc is reduced. For wavelengths longer than 4 $\mu$m, the light curve peak slowly decreases with the disc wall flux. The right panel of Figure \ref{fig:lcwave} shows the wavelength dependence of the magnitude difference. The wavelength for which the temperature variation effect is the largest depends on the temperature of the disc. For the fiducial model, the light curve modulations are strongest in the M' band. Hotter discs emit more strongly at shorter wavelength but there is also a stronger contribution from the central stars at those wavelengths so the peak remains smaller in K than M' except for high disc temperatures. The light curve variations also decrease more quickly with longer wavelength for hotter discs. For cooler discs, the largest light curve amplitude occurs at longer wavelengths. The temperature of the central stars also affects the wavelength dependence.  For hotter stars, the light shorter wavelengths have larger contributions from the stars and the size of the magnitude differences decreases more quickly. Cooler stars contribute less light so the variations from the disc are more prominent. For most low-mass binaries, the magnitude differences are largest between 3 $\mu$m and 10$\mu$m.

\begin{figure*}
\includegraphics[width=7.0in, trim= 0 0 0 0 ]{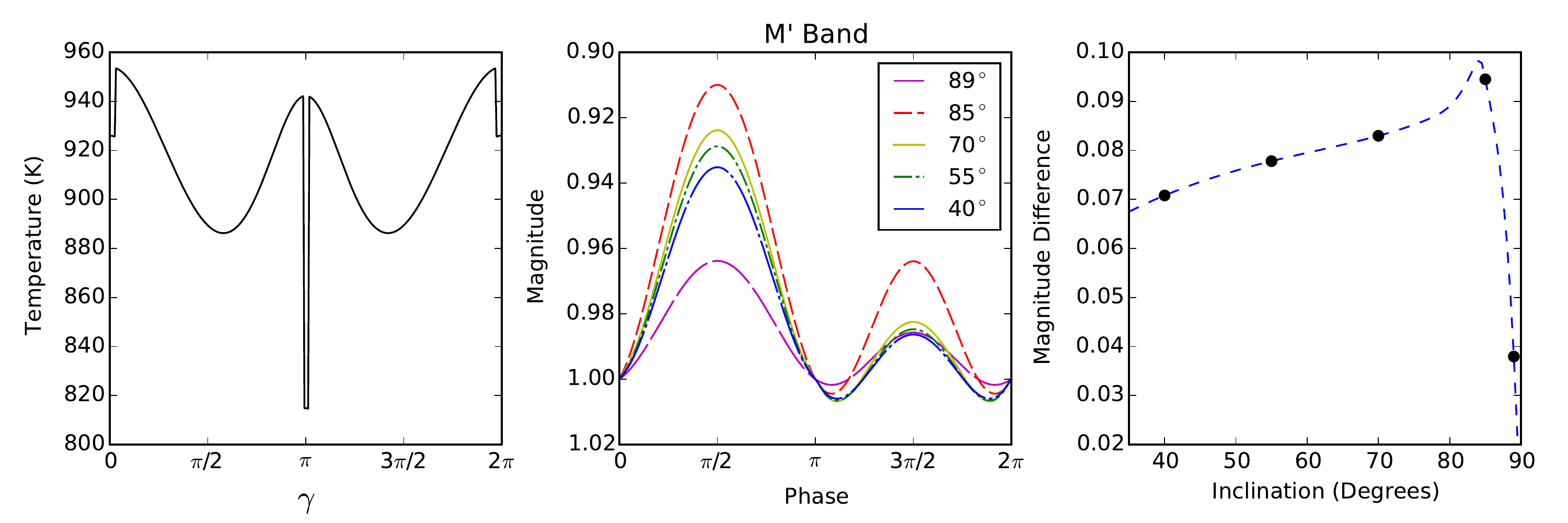}
\caption{Fiducial binary and circumbinary disc model but at different inclinations. The azimuthal temperature is shown in the left panel and is the same as in figure \ref{fig:lcwave}. The centre panel shows the M-band light curve but with $i=89^\circ$ in long dashed magenta, 85$^\circ$ in dashed red, 70$^\circ$ in solid yellow, 55$^\circ$ in dash-dot green, and 40$^\circ$ in long dash-dot blue. The right panel shows the M-band peak-to peak magnitude difference at different inclinations. The magnitude difference is largest at about 84$^\circ$.
}
\label{fig:lcinc}
\end{figure*} 

We varied the inclination of the system from fiducial model value of the 75$^\circ$ from 40$^\circ$ to 89$^\circ$. The inclination effects the size of the light curve variations but not the shape significantly, see Figure \ref{fig:lcinc}. Some small changes in the shape of the light curve occur from the edge of the disc obscuring itself at high inclinations but the most significant change is in the size of the magnitude difference. The largest magnitude difference of $\sim$0.1 mag for the fiducial model occurs at about 84$^\circ$. At higher inclinations, the amplitude drops very quickly because of the disc occultation. The inclination for the maximum amplitude depends on the geometry of the disc. Increasing the radius allows for higher inclinations without occultation while increasing the height of the disc requires lower inclinations. For the fiducial model, the binary star starts to eclipse the disc and the disc eclipse the binary for inclination over about 83$^\circ$ but those effects are neglected here. Those occultation would have significant effects on the light curve as they cause decreases in flux larger than variations predicted by this model. Light curve modulation from variations in the temperature profile are most likely to be observed at inclination slightly below the limit for the disc occulting the stars. However, the light curve modulations continue to prominent at moderately low inclinations. There is only a 0.02 mag reduction in the magnitude difference when the inclination is decreased from 70$^\circ$ to 40$^\circ$.
 
The light curve variations depends significantly on several parameters: stellar luminosity ratio ($L_{\star,1}/L_{\star,2}$), stellar mass ratio ($M_1/M_2$), orbital separation, and radius of the disc. We study each dependence individually using the fiducial model as a starting point.
 
 \subsection{Stellar Luminosity}

\begin{figure*}
\includegraphics[width=7.0in, trim= 0 0 0 0 ]{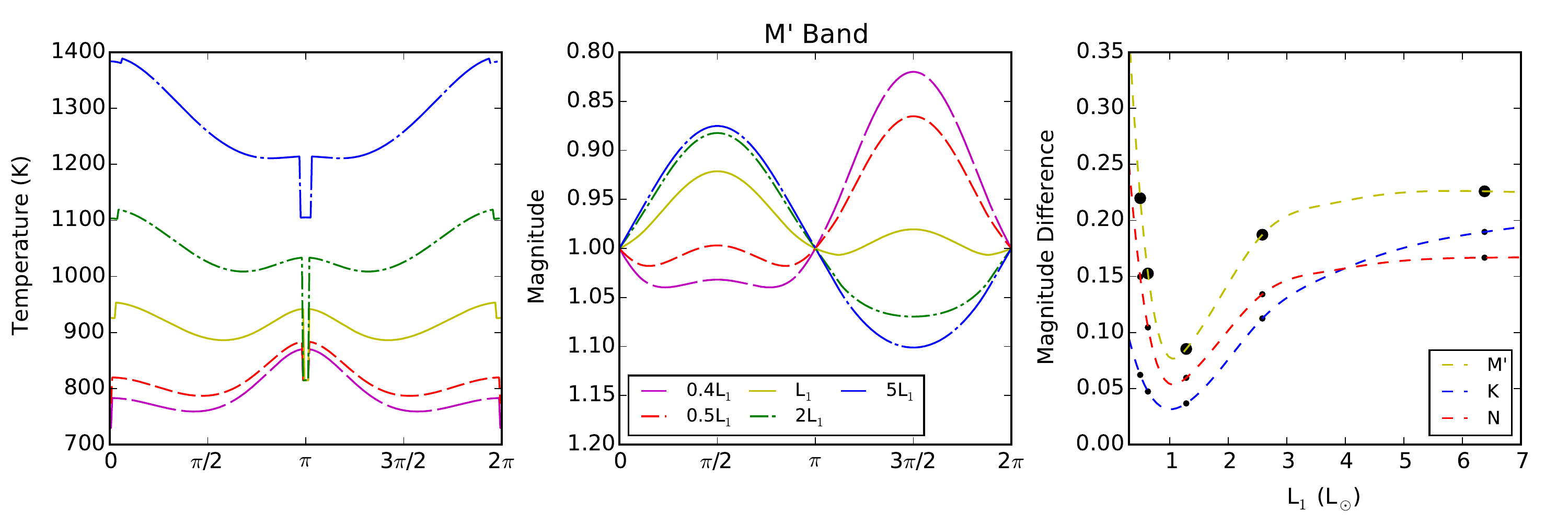}
\caption{Fiducial model but for primary stars of varying luminosities consistent with a 1 $M_\odot$ PMS star. Secondary star parameters kept constant. The left panels show the azimuthal temperature profiles and the centre panels are the associated M' band light curves for each temperature profile. Line styles and colours mark which temperature profile goes with which light curve.The lines are longed dashed magenta for $0.4L_1$, dashed red for $0.5L_1$, solid yellow $L_1$, dash-dot green for 2$L_1$, and long dash-dot blue for 5$L_1$ with $L_1$ that of the fiducial model. The right panels show the magnitude differences in the light curves  at three wavelengths, 2.2 (K), 4.68 (M') and 10.5 $\mu$m (N).
}
\label{fig:lclum}
\end{figure*} 

To study the effect of the stellar luminosity on the light curve, we kept all parameters from the fiducial model constant while varying the luminosity of the primary star. The luminosity was adjusted by changing the temperature and radius of star 1 in a way that is consistent with the evolutionary models of \cite{S2000}. This is equivalent to changing the age of star 1. The parameters of star 2 are kept constant. We also adjusted the height of the disc to account for the different average temperature of the disc wall and to keep the hydrostatic equilibrium condition satisfied. The luminosity was set to $0.4L_1$, $0.5L_1$, $L_1$, $2L_1$, and $5L_1$ where $L_1$ is the luminosity of star 1 in the fiducial model. The lowest value of luminosity used is equal to the lowest possible luminosity along the evolutionary track. Decreasing the luminosity is equivalent to increasing the age of the PMS star. The lowest luminosity occurs at about $\sim$10 Myr and the star is half the fiducial value at about 5 Myr. Decreasing the age to $\sim$0.8 Myr and $\sim$0.3 Myr increases the luminosity twice and five times the fiducial value. The primary is star 1 for every luminosity except for $0.4L_1$ where star 1 is slightly less luminous than star 2. 

The luminosity ratio, $L_1/L_2$, changes the depths of the dips in temperature due to stellar eclipses and changes the shape of the SED of the stars. The amplitude of variability in the azimuthal disc temperature increases as the stellar luminosity ratio departs from the fiducial model value (see left panel of Figure \ref{fig:lclum}). The temperature decreases with luminosity but the variability increases. 
The size of the variation reaches a minimum in both the temperature profile and the light curve when the primary to secondary luminosity ratio is equal to the mass ratio, $M_1/M_2$=2 which is slightly below the fiducial model luminosity ratio of 2.795. At this luminosity ratio, the temperature peaks are equal which leads to a sinusoidal light curve with a period twice that of the orbital period. The effect is very similar to when there are two identical stars.
Increasing the temperature variations by either decreasing or increasing the luminosity ratio from 2 causes an increase in the magnitude difference and the shape of the light curve becomes more sinusoidal as shown in the centre panel of figure \ref{fig:lclum}. The temperature dips from the shadow at $\gamma=\pi$ increases with temperature which causes a slightly larger dip in the light curve. The dependency of the magnitude difference on wavelength is similar to the fiducial model except at the largest luminosity ratio. In the M', and N bands, the magnitude differences are are similar and the difference in K band is much lower except for larger luminosities. At $5L_1$, the magnitude difference is no longer the smallest in K but is the smallest instead at a much longer wavelength in N band. For more luminous stars, the disc wall temperature is much hotter so that the largest magnitude differences occur at shorter wavelengths in the near-IR. Despite the unrealistic temperature adjustments to the model, we expected the behaviour of the light curves to be very similar if more the luminosity was adjusted in a way more consistent with stellar models.

 \subsection{Stellar Mass}
 
 \begin{figure*}
\includegraphics[width=7.0in, trim= 0 0 0 0 ]{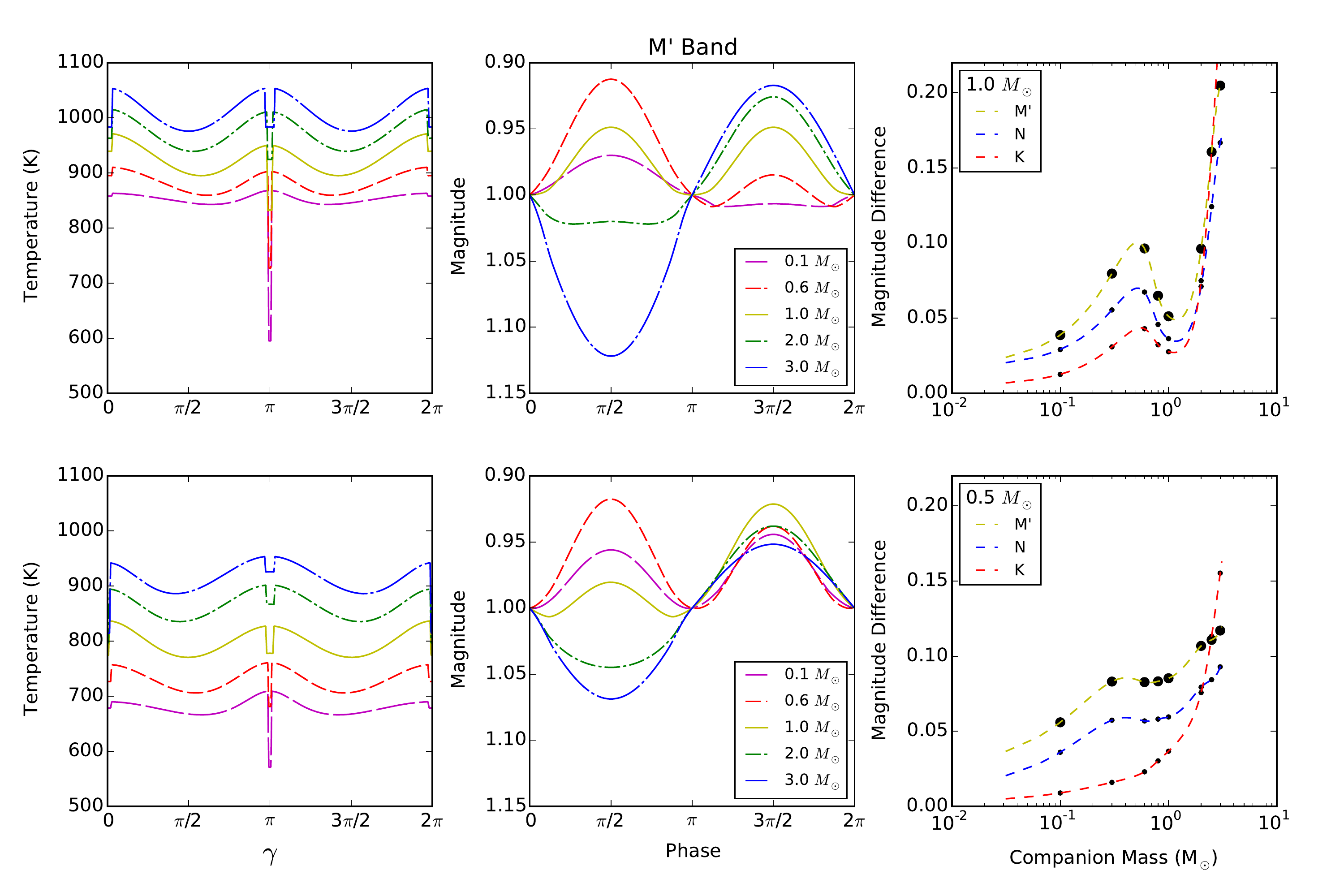}
\caption{Fiducial model but with star 2 varying mass and varying luminosity consistent with a 2 Myr PMS star using evolution models from \citet{S2000}. Star 1 parameters are kept constant. The left and the centre panels are the temperature profiles and the light curves, respectively. Matching line styles and colours mark temperature curves with its associated light curve. The companion masses are 0.1 (long dashed magenta), 0.6 (dashed red), 1.0 (solid yellow), 2.0 (dash-dot green), and 3.0 $M_\odot$ (long dash-dot blue). The right panels are the magnitude differences as a function of star 2 mass in K (red), M' (yellow) and N band (blue). In the top panels, star 1 mass is 1 $M_\odot$. In he bottom panels, star 1 mass is 0.5 $M_\odot$. Star 1 is not always the primary so for a larger mass star 2, the peak of the light curve is at $\phi$=3$\pi$/2 instead of $\phi$=$\pi$/2.
}
\label{fig:lcmass}
\end{figure*} 
 
\begin{figure*}
\includegraphics[width=7.0in]{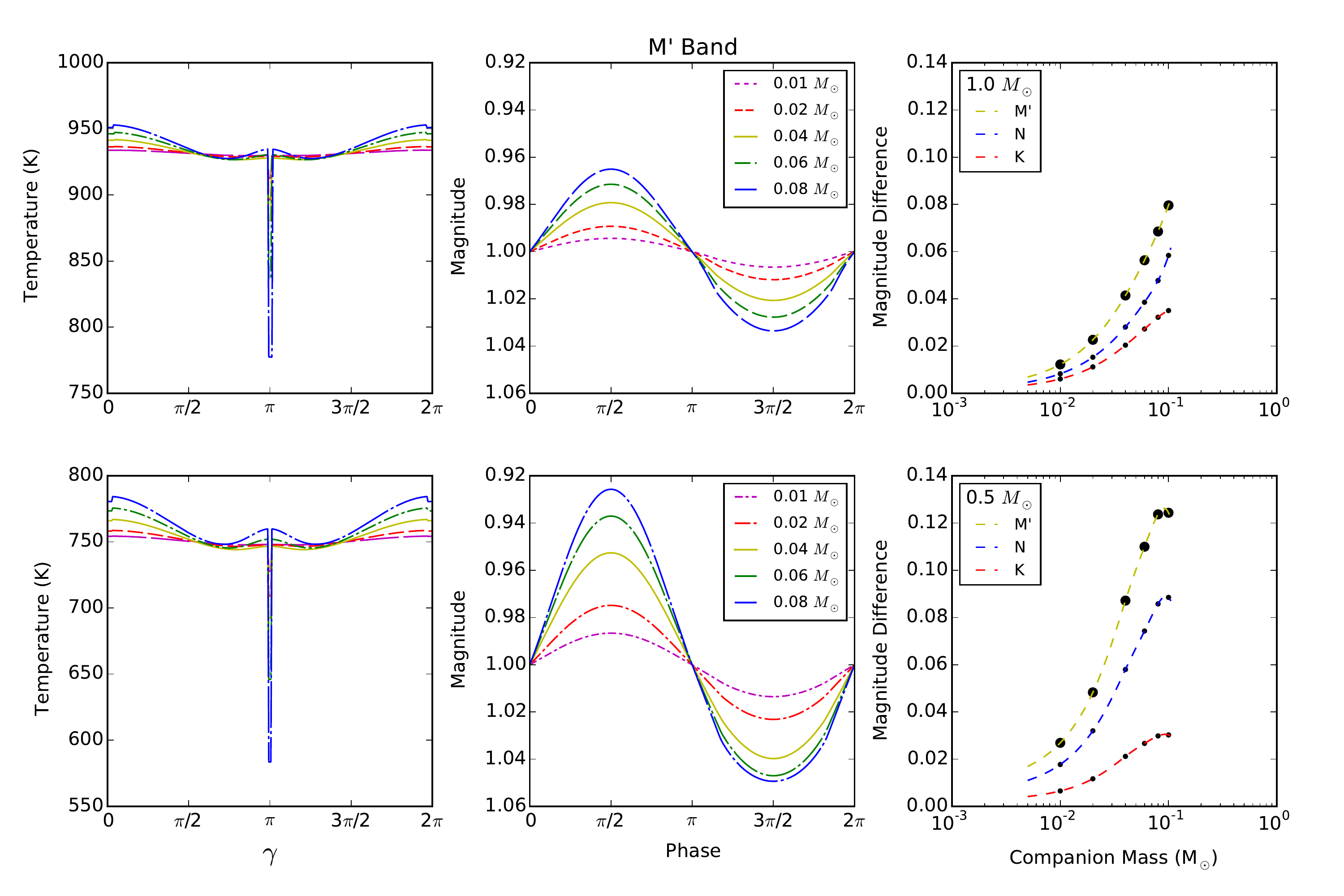}
\caption{Fiducial model orbital parameters with star 2 varying mass and varying luminosity consistent with brown dwarfs at age 1 Myr using models from \citet{B2003} and star 1 parameters are kept constant as a 1 Myr PMS star. In the top panels, star 1 mass is 1 $M_\odot$. In he bottom panels, star 1 mass is 0.5 $M_\odot$. The left and the centre panels are the temperature profiles and the light curves, respectively. Matching line styles and colours mark temperature curves with its associated light curve. The companion masses are 0.01 (long dashed magenta), 0.02 (dashed red), 0.04 (solid yellow), 0.06 (dash-dot green), and 0.08 $M_\odot$ (long dash-dot blue). The right panels are the magnitude differences as a function of star 2 mass in K (red), M' (yellow) and N band (blue).
\label{fig:brown}
}
\end{figure*} 
 
In this section, we vary the mass of one of the stars while keeping the stellar parameters consistent with \citeauthor{S2000}'s (\citeyear{S2000}) stellar evolution models of 2 Myr old PMS star. First we keep one star the same at 1 $M_\odot$  while varying the companion over a range of masses (see top panels of Figure \ref{fig:lcmass}) and then keep one star at 0.5 $M_\odot$ while varying the companion over the same mass range (see bottom panels of same figure). The companion masses range from 0.1 to 3.0 $M_\odot$. The orbital separation remains the same as the fiducial model at 30 $R_\odot$ for each pair and the inclination at 75$^\circ$. 

The stellar mass ratio sets the semi-major axes of the central stars. As the mass of the star increases, it moves closer to the centre of mass and with a smaller orbit, the size of the distance oscillations shrink compared to the distance to the disc wall. However, the luminosity increases with mass so the temperature of the disc wall increases along with its variations. The light curve becomes more sinusoidal like the previous section. With decreasing mass of star 2, the orbit of star 1 shrinks. With smaller distance oscillations and less stellar irradiation, the disc wall cools and the azimuthal variations in temperature become smaller.  In the light curve, the decrease in temperature variations leads to smaller magnitude difference and a less sinusoidal curve.
As the primary moves inward with increasing mass ratio, the secondary moves closer to the disc wall. When close enough to the wall, the secondary becomes the dominant source of temperature variation and the peak of the light curves changes from $\pi/2$ to $3\pi/2$. In the bottom centre panel of figure \ref{fig:lcmass}, this occurs between the red and yellow lines which are for star 2 masses of 0.1 and 0.6 $M_\odot$ and star 1 mass of 0.5 $M_\odot$. This does not occur for the 1.0 $M_\odot$ star.

The largest difference occurs for companions with masses larger than 2 $M_\odot$ for both masses but the magnitude differences are bigger for the 1 $M_\odot$ star.  There is a local minimum in the magnitude difference when the stars are identical that is more pronounced for the 1 $M_\odot$ star, see the left panels of Figure \ref{fig:lcmass}. In the upper left panel, there is a small peak in magnitude difference at about 0.5 $M_\odot$ companion mass for the 1 $M_\odot$ star and then the difference steadily decreases.  For the lower masses, the companion has too low of a luminosity to induce strong variations in the temperature despite being closer to the disc and the primary's orbit becomes too small to significantly vary the distance between it and the wall. The 0.5 $M_\odot$ does not have a sharp decrease in magnitude difference for an identical PMS companion and stays around 0.8 mag in M' band within the range 0.3 to 1.0 $M_\odot$. 

To explore masses lower than 0.1 $M_\odot$, we change the age of the binary star to 1 Myr and use brown dwarf evolutionary models from \cite{B2003} for the companions. The primary is again a 1 $M_\odot$ PMS star and then a 0.5 $M_\odot$ PMS star. The mass of the secondary is varied from 0.01 to 0.1 $M_\odot$ which covers the brown dwarf companion regime. The temperature profiles, light curves and magnitude differences are shown in the left, centre and right panels in Figure \ref{fig:brown}, respectively. For both primary masses, the magnitude difference decreases with decreasing secondary mass but the decline is quicker for the 0.5 $M_\odot$ star. As the system ages, the magnitude difference will remain about the same as the brown dwarf cools and retracts since the PMS star is the source of most of the light curve variations. The dip in the temperature profile and the light curve decrease as the eclipse shadow region decreases but most of the light curve variation is caused by the primary unevenly irradiating the disk so the effect of decreasing eclipse shadow is small. Low mass companions cause more variability in the light curve for low mass primaries.

\subsection{Binary Separation}

 \begin{figure*}
\includegraphics[width=7.0in, trim= 0 0 0 0 ]{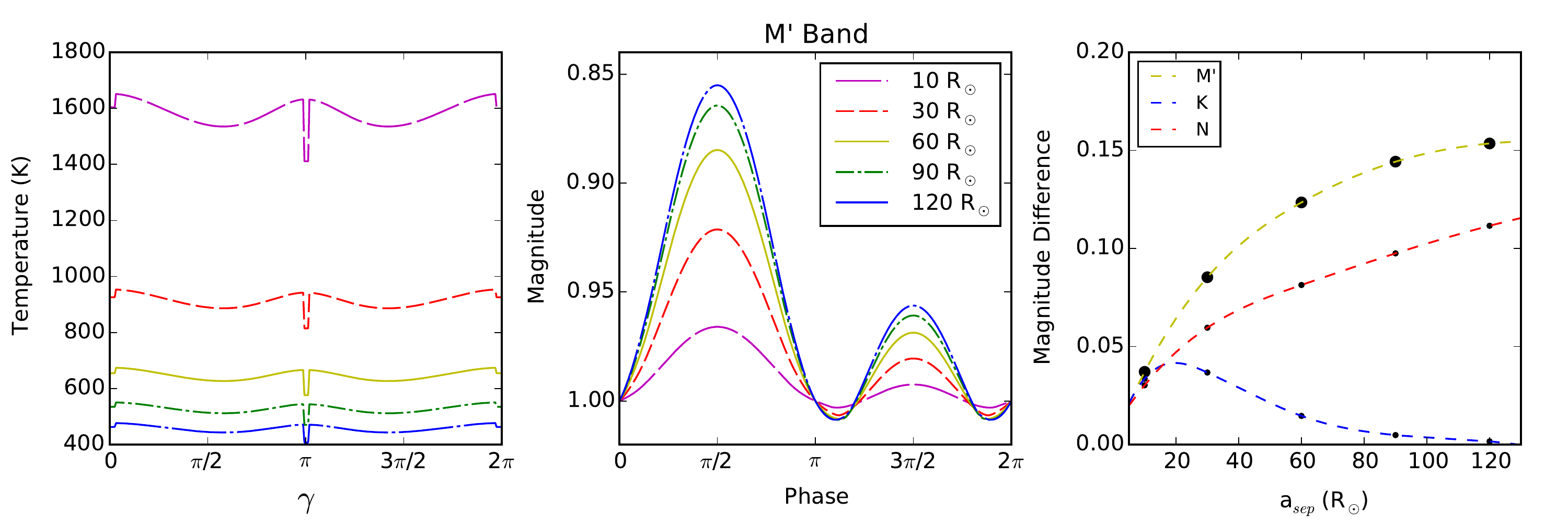}
\caption{Fiducial model but with changing orbital separation. The left and centre panels are the temperature profiles and the light curves, respectively, with matching line styles matching profiles with light curves. The separations used are 10 $R_\odot$ (0.046 AU), 30 $R_\odot$ (0.14 AU), 60 $R_\odot$ (0.28 AU), 90 $R_\odot$ (0.42 AU), and 120 $R_\odot$ (0.56 AU) in long dashed magenta, dashed red, solid yellow, dash-dot green, and long dash-dot blue. The fiducial model is the dashed red line. The right panel is the magnitude difference versus orbital separation in K (blue), M' (yellow), and N band (red).
}
\label{fig:lcsep}
\end{figure*} 

To study the dependency on the binary separation, we vary the separation while keeping most of the fiducial parameters constant. Since the disc wall radius is defined in relation to the orbital separation, varying the separation requires adjusting the radius of inner disc wall. The height also increases with the disc radius even though the disc wall temperature decreases. The stellar properties are kept to the values in the fiducial model. In Figure \ref{fig:lcsep}, we vary the orbital separation from 10 $R_\odot$ to 120 $R_\odot$. As the orbital separation increases, the temperature of the disc wall decreases and the variations in temperature decrease proportionally. The shape of the light curve remains the same with varying orbital separation. Only the peaks change in size, increasing with decreasing temperature.  The magnitude difference is very colour sensitive. The amplitude of the light curve variations increases with increasing separation in M' and N band but in K, the amplitude peaks about 0.04 mag around 20$ R_\odot$. This occurs because blackbodies peak in the K band at about 1300 K and the maximum variation occurs near the peak wavelength of the blackbody. For the M' band, the temperature is 600 K so the magnitude difference starts to decrease around 120 $R_\odot$. For N, the temperature for the peak wavelength to be that band is about 300 K so the magnitude difference does not reach its maximum in that band in the range we studied. Binaries with longer periods have larger light curve modulation in the mid-IR.
 
 \subsection{Disc Radius}
 
\begin{figure*}
\includegraphics[width=7.0in, trim= 0 0 0 0 ]{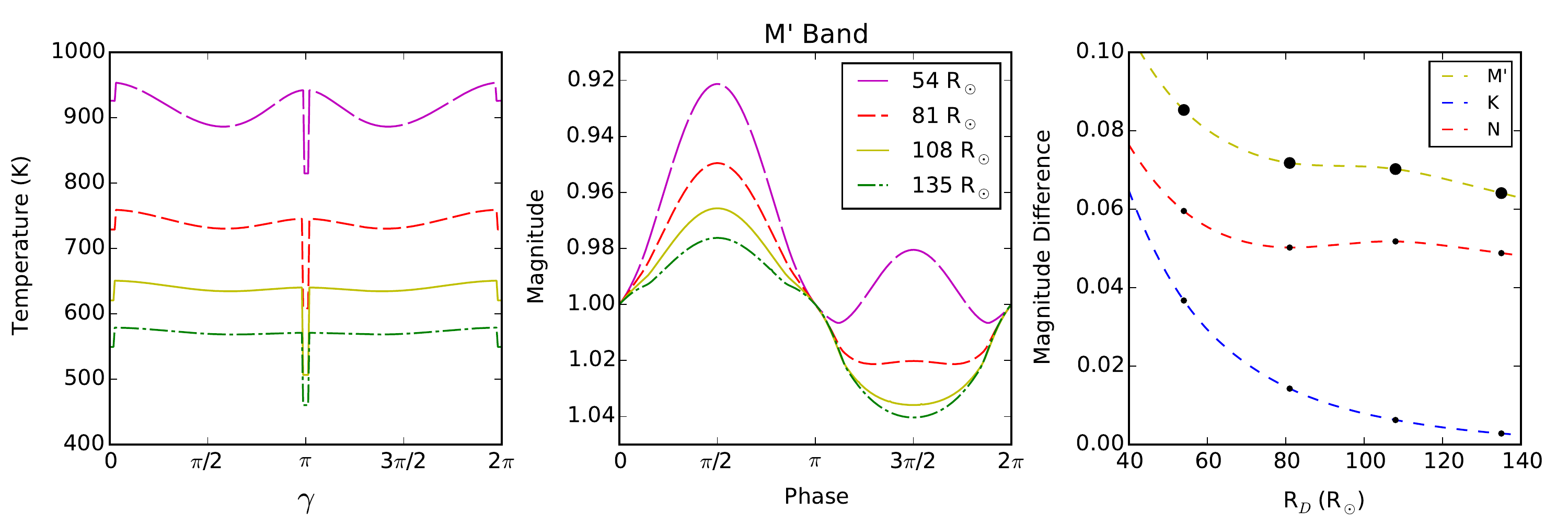}
\caption{Fiducial model but with changing radius of the disc wall. The left and centre panels are the temperature profiles and light curves with matching line styles and colours. The disc radii used are 54 $R_\odot$ ($R_{D,\text{fid}}$ = 0.25 AU), 81 $R_\odot$ (1.5$R_{D,\text{fid}}$ = 0.38 AU), 108 $R_\odot$ (2$R_{D,\text{fid}}$ = 0.5 AU), 135 $R_\odot$ (2.5$R_{D,\text{fid}}$ = 0.63 AU) in long dashed magenta, dashed red, solid yellow, and dash-dot green. The right panel is the magnitude difference in K (blue), M' (yellow), and N band (red).
}
\label{fig:lcdrad}
\end{figure*} 

The last parameter we explore is the circumbinary disc radius. In the fiducial model, the disc radius is set to the minimum dynamically stable orbit, $R_D=1.8 a_\text{sep}=54\ R_\odot$. The disc wall can be much farther from this point because of processes such as dust sublimation or photoevaporation. A strong photoevaporation wind can slightly increase the disc wall radius \citep{A2012}. We increase the radius by a factor of 1.5, 2, and 2.5. Generally, the temperature profiles flatten with larger disc radii with the dips in temperature from the stellar eclipse shadows decreasing slowest (see left panel of Figure \ref{fig:lcdrad}). Unlike the orbital separation case, the light curve shape changes significantly with increasing disc radius. Both peaks in the light curve shrink and the smaller peak becomes a dip (see centre panel Figure \ref{fig:lcdrad}) and resembles dips caused by extinction. The peaks in the temperature curve flatten which leaves only the temperature dips from the stellar eclipses as the source of variation. At all wavelengths studied, the magnitude differences decrease with increasing wall radius with the longer wavelengths decreasing more slowly. The light curve variations in K reach 0 at about $R_D=140 R_\odot$. The differences between light curve variations between M' and N are small ($\lesssim$2) in the radius range.

Dust sublimation occurs at approximately 1500 K. For the fiducial model, the disc is already well below this temperature but for a closer binary, the temperature can be above this limit if the disc wall is at the minimum stable radius, see 10 $R_\odot$ separation profile in Figure \ref{fig:lcsep}. For binaries closer together than our fiducial model, there is an initial increase in the light curve amplitude as the temperature of the disc decreases with increasing radius before the amplitude decreases like in Figure \ref{fig:lcdrad}. This initial increase is from the peak blackbody wavelength increasing to the IR bands. Generally, discs with wider clearings have much smaller variations.

 \subsection{Application to CoKu Tau/4}\label{sec:coku}

\begin{table}
\begin{center}
\vbox to 55mm{\vfil
\caption{\large Coku Tau/4 Parameters}
\begin{tabular}{@{}llc}
\hline
& & Ref. \\
\hline
$M_1$  & 0.6 $M_\odot$  & 1 \\
$M_2$   & 0.5 $M_\odot$  & 1     \\
$R_1$ & 1.261 $R_\odot$ & 1\\
$R_2$  & 1.178 $R_\odot$ & 1  \\
$T_{\text{eff},1}$ & 4027 K  & 1\\
$T_{\text{eff},2}$ & 3894 K  & 1\\
$a_\text{sep}$ &1598 $R_\odot$ $\approx$7.43 AU & 2\\
$R_D$ & 2877 $R_\odot$ $\approx$13 AU & 2\\
$i$ & 60 deg & 2 \\
$H/R_D$ & 0.11\\
\hline
\end{tabular}
{\\Parameters for model of Coku Tau/4 system \\
References: 1- \cite{S2000} ; 2- \cite{N2010}\\
\label{tab:coku}}
\vfil}
\end{center}
\end{table}

\begin{figure*}
\includegraphics[width=7.0in, trim= 0 0 0 0 ]{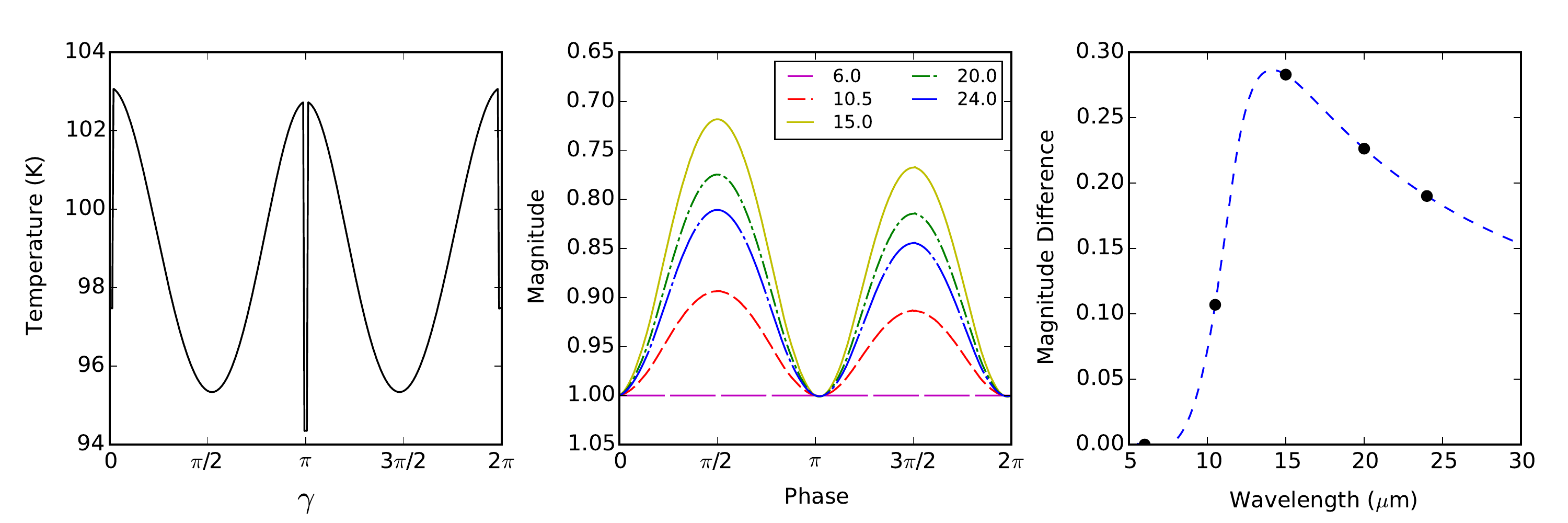}
\caption{Light curve model for the CoKu Tau/4 at different wavelengths. Left panel shows the azimuthal temperature profile. The middle panel is the light curves at 6.0 $\mu$m (long dashed magenta), 10.5 $\mu$m (dashed red), 15.0 $\mu$m (solid yellow), 20.0 $\mu$m (dash-dot green), and 24.0 $\mu$m (long dash-dot blue). The right panel is the magnitude difference vs wavelength.
}
\label{fig:coku}
\end{figure*} 

We apply our model to the known binary system CoKu Tau/4 that \cite{N2010} predicted would have light curve variations from the temperature oscillations. The Coku Tau/4 system is binary star comprising similar mass PMS companions and a circumbinary disc. To apply our model to this system, we use the same stellar parameters used by \cite{N2010} for their SED model. The age of the stars is set to 4 Myr and the properties of the stars are taken from evolution models by \cite{S2000}. We use $a_\text{sep}=1598\ R_\odot=7.43$ AU, an inclination of 60$^\circ$ and $R_D=1.8 a_\text{sep}$ which are the same values \cite{N2010} used for their fiducial model. We use a slightly different value for the height of the disc wall. Our wall height is set by hydrostatic equilibrium like our fiducial model which is slightly smaller than the height used by \cite{N2010}. All the model parameters are list in table \ref{tab:coku}. We computed the light curve of the system within the wavelength range of 5 $\mu$m to 30 $\mu$m (see Figure \ref{fig:coku}). 

The light curve has two peaks of similar height since the stars have very similar luminosities and masses. We find no variation in the light curved for wavelengths less than 9 $\mu$m but predict $\sim$10\% variation at 10 $\mu$m and about a 20\% amplitude variation for wavelengths larger than 15 $\mu$m. \cite{N2010} predicted $\sim$10\% variation in the mid-IR SED of the CoKu Tau/4 system over half it's orbital period which is in good agreement with our model. Despite the strong 10$\mu$m silicate feature in the spectra of their model, our model's prediction of size of the variations is within about a factor of a couple. The largest difference between the models is our model predicts a peak magnitude difference of 0.28 mag at 15 $\mu$m that does not appear in their spectra. Their model predicts an approximately 0.1 mag difference over the entire range studied with little dependence on the wavelength (see their Figure 11). The dust feature extends to about 15 $\mu$m and likely diminishes the temperature oscillation effects near our predicted peak. We expect the inclusion of dust features to change the wavelength dependence of the amplitude of the light curve variations but should not significantly change the shape.
 
 \subsection{Other Sources}
 
 \begin{table}
\begin{center}
\vbox to 40mm{\vfil
\caption{\large IR Variability Sources}
\begin{tabular}{@{}ll}
Source & $\Delta$M'  \\
\hline
Disc Wall Temperature Modulations  & 0.08   \\
Cool Spots$^1$ & 0.2$^a$/0.07$^b$       \\
Hot Spots$^1$ & 0.4$^a$/0.1$^b$   \\
Disc Occultation$^c$  & 0.1   \\
Stellar Eclipse$^c$ & 0.5  \\
\hline
\end{tabular}
{\\Notes: a- Spot covers 30\% of star; b- Spot covers 10\%; c- Refers to primary eclipse
\\References: 1- \cite{CHS2001} \\
\label{tab:compare}}
\vfil}
\end{center}
\end{table}
 
There are multiple sources of IR variability that likely occur simultaneously \cite{CSB2014}. Variability in the near and mid-IR is most commonly attributed to star spots, accretion, and disc extinction \cite[e.g.][]{WRA2013,PPWG2014}. Accretion variations should not be periodic and the time-scale depends on the mechanism causing the variation \cite[e.g.][]{CHS2001}. Disc extinction can be periodic or occur on long time-scales. Spots are periodic on the timescale of stellar rotation.
 
We listed periodic sources of IR variability in Table \ref{tab:compare} with estimates of the magnitude variance. The cool and hot spots were calculated with a simple spot model \cite{CHS2001} using a spot that is half and twice the temperature of the fiducial primary, respectively, covering 30\% of the star and covering 10\% of the star. Cool spots variations are about the same size as the variations from the temperature modulations but the spot light curve has a sinusoidal shape instead of double peaks of this model. Hot spot are a factor of a few larger and the light curve, if the spots are long lived, is the same as the cool spots. The stellar eclipse calculated is the primary eclipse of a binary set the fiducial model stellar parameters and no circumbinary disc. The disc occultation is also calculated with the fiducial model stellar parameters and orbital separation. Table \ref{tab:compare} list the size of the primary stellar eclipse and the size of the dip if the disc completely occults the primary. The disc wall temperature modulations are slightly smaller than the other sources of  periodic variations but the light curve shape is unique.
 
 \section{Conclusion}\label{sec:con}
 
Temperature oscillations along the inner wall of a circumbinary disc arise from the phase dependence of the distance between each star in a binary and positions along the disc wall.  These temperature variations can give  significant periodic modulation in a photometric infrared light curve.  We compute a magnitude difference greater than 5 per cent for wavelengths longer than $\sim3\ \mu$m  in a light curve from a pre-main sequence solar mass binary with mass ratio 2, period of 15 days and inclination of $75^\circ$ embedded in a circumbinary disc with a cylindrical inner edge. The maximum magnitude difference was about 10 per cent at $\sim4\ \mu$m. We find that the light curve exhibits a non-sinusoidal shape consisting of two peaks of different heights. One star can block the light of the other, shadowing the disc wall, however, these shadows do not cause significant variations in the integrated light curve. The size and shape of the light curve modulations are sensitive to the primary to secondary luminosity ratio, the mass ratio, binary orbital separation and the radius of the circumbinary disc wall. Larger luminosity or mass ratios can increase the size of the estimated photometric variations by a factor of a few and shape of the light curve becomes more sinusoidal. Increasing the orbital separation increases the size of the light curve variations at wavelengths longer than $\sim2.5\ \mu$m and does not affect the light curve shape. Increasing the disc radius decreases the magnitude difference and changes the shape from two peaks to a small peak with a large dip. Light curve modulations from temperature oscillations are strongest for two dissimilar mass stars with orbital periods of the order of a couple weeks at wavelengths over $\sim3\ \mu$m. 

Because most young stars are born as binaries, infrared photometric surveys of young stars should find a class of periodically varying objects that are associated with emission from circumbinary disc walls.  The shapes of the light curves we have illustrated here may aid in identifying periodic modulations associated with circumbinary material. Significant amplitude variations could be present even when a binary is oriented with an inclination as low as $40^\circ $ to the line of light, so a binary need not be eclipsing to be identified if a circumbinary disc is present. While we have restricted this initial study to cylindrical shaped disc walls and circular binary orbits,  more complex models might make it possible to constrain the shape of a circumbinary disc edge. The photometric level of accuracy required to detect modulations due to a circumbinary disc is not so exacting as to require a space observatory. In comparison reflection, and relativistic beaming require a much higher level of photometric accuracy. Beaming studies have discovered hundreds of non-eclipsing binary systems \citep{TFM2015}. Hence infrared photometric surveys \citep[e.g.][]{CSB2014} may be quite sensitive to previously undetected circumbinary discs.

Our model does not include any radiative transfer and emission from disc beyond the circumbinary disc wall was ignored. Depending on how the fluctuations in temperature along the disc wall propagate outward, radiative transfer could strengthen or diminish the amplitude of light curve variations. \cite{Demidova2013} found that the radius to which illumination asymmetries created by shadows of small mass companion can be observed depends on the viscosity of the disc. Asymmetries are more quickly reduced in low viscosity or "cold" discs but such asymmetries only extend out a small fraction of the total disc. However, if the disc emissions are dominated by the wall, then our model gives a good approximation for the shape of the light curves and the size-scale of flux variations.

Thermal time-scales were also ignored in our model. If the time for thermal equilibrium is much shorter than the orbital time-scale, then neglecting heating and cooling time-scale is a good approximation. However if the time-scale is longer or comparable to the angular velocity of the disc wall, this temperature oscillations would smear out and there would be no light curve variations. If there is a slight lag in the thermal equilibrium, this could create an offset between the light curve peaks and the stellar eclipses.

The thin cylindrical shape for the inner edge of the circumbinary disc is a unrealistic simplification. An eccentric disc wall would impact the temperature along the edge and the observed flux modulations. The strength of the effects from an eccentric disc wall should depend on the orientation of the ellipse and its eccentricity. The more distant regions of the wall would be much cooler and vary less from star to disc distance oscillations. Eccentric binary orbits would also affect the flux modulations and depend on the orientation of the orbits since some regions of the disc wall would vary more in temperature from the stars' movements.

\section*{Acknowledgements}
We thank Eric Mamajek for reading this paper and providing helpful comments. 

\bibliography{diskmodel}

\begin{thebibliography}{}

\bibitem[\protect\citeauthoryear{{Alexander}}{{Alexander}}{2012}]{A2012}
{Alexander} R.,  2012, \apjl, 757, L29

\bibitem[\protect\citeauthoryear{{Artymowicz} \& {Lubow}}{{Artymowicz} \&
  {Lubow}}{1994}]{AL1994}
{Artymowicz} P.,  {Lubow} S.~H.,  1994, \apj, 421, 651

\bibitem[\protect\citeauthoryear{{Baraffe}, {Chabrier}, {Barman}, {Allard} \&
  {Hauschildt}}{{Baraffe} et~al.}{2003}]{B2003}
{Baraffe} I.,  {Chabrier} G.,  {Barman} T.~S.,  {Allard} F.,    {Hauschildt}
  P.~H.,  2003, \aap, 402, 701

\bibitem[\protect\citeauthoryear{{Bell}, {Naylor}, {Mayne}, {Jeffries} \&
  {Littlefair}}{{Bell} et~al.}{2013}]{B2013}
{Bell} C.~P.~M.,  {Naylor} T.,  {Mayne} N.~J.,  {Jeffries} R.~D.,
  {Littlefair} S.~P.,  2013, \mnras, 434, 806

\bibitem[\protect\citeauthoryear{{Biller}, {Lacour}, {Juh{\'a}sz}, {Benisty},
  {Chauvin}, {Olofsson}, {Pott}, {M{\"u}ller}, {Sicilia-Aguilar}, {Bonnefoy},
  {Tuthill}, {Thebault}, {Henning} \& {Crida}}{{Biller}
  et~al.}{2012}]{Biller2012}
{Biller} B.,  {Lacour} S.,  {Juh{\'a}sz} A.,  {Benisty} M.,  {Chauvin} G.,
  {Olofsson} J.,  {Pott} J.-U.,  {M{\"u}ller} A.,  {Sicilia-Aguilar} A.,
  {Bonnefoy} M.,  {Tuthill} P.,  {Thebault} P.,  {Henning} T.,    {Crida} A.,
  2012, \apjl, 753, L38

\bibitem[\protect\citeauthoryear{{Carpenter}, {Hillenbrand} \&
  {Skrutskie}}{{Carpenter} et~al.}{2001}]{CHS2001}
{Carpenter} J.~M.,  {Hillenbrand} L.~A.,    {Skrutskie} M.~F.,  2001, \aj, 121,
  3160

\bibitem[\protect\citeauthoryear{{Clanton}}{{Clanton}}{2013}]{C2013}
{Clanton} C.,  2013, \apjl, 768, L15

\bibitem[\protect\citeauthoryear{{Cody}, {Stauffer}, {Baglin}, {Micela},
  {Rebull}, {Flaccomio}, {Morales-Calder{\'o}n}, {Aigrain}, {Bouvier},
  {Hillenbrand}, {Gutermuth}, {Song}, {Turner}, {Alencar} \& et al}{{Cody}
  et~al.}{2014}]{CSB2014}
{Cody} A.~M.,  {Stauffer} J.,  {Baglin} A.,  {Micela} G.,  {Rebull} L.~M.,
  {Flaccomio} E.,  {Morales-Calder{\'o}n} M.,  {Aigrain} S.,  {Bouvier} J.,
  {Hillenbrand} L.~A.,  {Gutermuth} R.,  {Song} I.,  {Turner} N.,  {Alencar}
  S.~H.~P.,    et al 2014, \aj, 147, 82

\bibitem[\protect\citeauthoryear{{D'Alessio}, {Cant{\"o}}, {Calvet} \&
  {Lizano}}{{D'Alessio} et~al.}{1998}]{DCL1998}
{D'Alessio} P.,  {Cant{\"o}} J.,  {Calvet} N.,    {Lizano} S.,  1998, \apj,
  500, 411

\bibitem[\protect\citeauthoryear{{D'Alessio}, {Hartmann}, {Calvet},
  {Franco-Hern{\'a}ndez}, {Forrest}, {Sargent}, {Furlan}, {Uchida}, {Green},
  {Watson}, {Chen}, {Kemper}, {Sloan} \& {Najita}}{{D'Alessio}
  et~al.}{2005}]{DHC2005}
{D'Alessio} P.,  {Hartmann} L.,  {Calvet} N.,  {Franco-Hern{\'a}ndez} R.,
  {Forrest} W.~J.,  {Sargent} B.,  {Furlan} E.,  {Uchida} K.,  {Green} J.~D.,
  {Watson} D.~M.,  {Chen} C.~H.,  {Kemper} F.,  {Sloan} G.~C.,    {Najita} J.,
  2005, \apj, 621, 461

\bibitem[\protect\citeauthoryear{{Demidova}, {Grinin} \&
  {Sotnikova}}{{Demidova} et~al.}{2013}]{Demidova2013}
{Demidova} T.~V.,  {Grinin} V.~P.,    {Sotnikova} N.~Y.,  2013, Astronomy
  Letters, 39, 26

\bibitem[\protect\citeauthoryear{{Duch\^ene} \& {Kraus}}{{Duch\^ene} \&
  {Kraus}}{2013}]{DK2013}
{Duch\^ene} G.,  {Kraus} A.,  2013, \araa, 51, 269

\bibitem[\protect\citeauthoryear{{Fernandez} \& {Eiroa}}{{Fernandez} \&
  {Eiroa}}{1996}]{FE1996}
{Fernandez} M.,  {Eiroa} C.,  1996, \aap, 310, 143

\bibitem[\protect\citeauthoryear{{Gillen}, {Aigrain}, {McQuillan}, {Bouvier},
  {Hodgkin}, {Alencar}, {Terquem}, {Southworth}, {Gibson}, {Cody}, {Lendl},
  {Morales-Calder{\'o}n}, {Favata}, {Stauffer} \& {Micela}}{{Gillen}
  et~al.}{2014}]{Gillen2014}
{Gillen} E.,  {Aigrain} S.,  {McQuillan} A.,  {Bouvier} J.,  {Hodgkin} S.,
  {Alencar} S.~H.~P.,  {Terquem} C.,  {Southworth} J.,  {Gibson} N.~P.,  {Cody}
  A.,  {Lendl} M.,  {Morales-Calder{\'o}n} M.,  {Favata} F.,  {Stauffer} J.,
  {Micela} G.,  2014, \aap, 562, A50

\bibitem[\protect\citeauthoryear{{Ireland} \& {Kraus}}{{Ireland} \&
  {Kraus}}{2008}]{IK2008}
{Ireland} M.,  {Kraus} A.,  2008, \apjl, 678, L59

\bibitem[\protect\citeauthoryear{{Jensen}, {Dhital}, {Stassun}, {Patience},
  {Herbst}, {Walter}, {Simon} \& {Basri}}{{Jensen} et~al.}{2007}]{Jensen2007}
{Jensen} E.~L.~N.,  {Dhital} S.,  {Stassun} K.~G.,  {Patience} J.,  {Herbst}
  W.,  {Walter} F.~M.,  {Simon} M.,    {Basri} G.,  2007, \aj, 134, 241

\bibitem[\protect\citeauthoryear{{Ma{\'{\i}}z Apell{\'a}niz}, {Negueruela},
  {Barb{\'a}}, {Walborn}, {Pellerin}, {Sim{\'o}n-D{\'{\i}}az}, {Sota}, {Marco},
  {Alonso-Santiago}, {Sanchez Bermudez}, {Gamen} \& {Lorenzo}}{{Ma{\'{\i}}z
  Apell{\'a}niz} et~al.}{2015}]{Maiz2015}
{Ma{\'{\i}}z Apell{\'a}niz} J.,  {Negueruela} I.,  {Barb{\'a}} R.~H.,
  {Walborn} N.~R.,  {Pellerin} A.,  {Sim{\'o}n-D{\'{\i}}az} S.,  {Sota} A.,
  {Marco} A.,  {Alonso-Santiago} J.,  {Sanchez Bermudez} J.,  {Gamen} R.~C.,
  {Lorenzo} J.,  2015, ArXiv e-prints

\bibitem[\protect\citeauthoryear{{Martin} \& {Triaud}}{{Martin} \&
  {Triaud}}{2015}]{MT2015}
{Martin} D.~V.,  {Triaud} A.~H.~M.~J.,  2015, \mnras, 449, 781

\bibitem[\protect\citeauthoryear{{Morales-Calder{\'o}n}, {Stauffer},
  {Hillenbrand}, {Gutermuth}, {Song}, {Rebull}, {Plavchan}, {Carpenter},
  {Whitney}, {Covey}, {Alves de Oliveira}, {Winston}, {McCaughrean} \& et
  al}{{Morales-Calder{\'o}n} et~al.}{2011}]{MSH2011}
{Morales-Calder{\'o}n} M.,  {Stauffer} J.~R.,  {Hillenbrand} L.~A.,
  {Gutermuth} R.,  {Song} I.,  {Rebull} L.~M.,  {Plavchan} P.,  {Carpenter}
  J.~M.,  {Whitney} B.~A.,  {Covey} K.,  {Alves de Oliveira} C.,  {Winston} E.,
   {McCaughrean} M.~J.,    et al 2011, \apj, 733, 50

\bibitem[\protect\citeauthoryear{{Nagel}, {D'Alessio}, {Calvet}, {Espaillat},
  {Sargent}, {Hern{\'a}ndez} \& {Forrest}}{{Nagel} et~al.}{2010}]{N2010}
{Nagel} E.,  {D'Alessio} P.,  {Calvet} N.,  {Espaillat} C.,  {Sargent} B.,
  {Hern{\'a}ndez} J.,    {Forrest} W.~J.,  2010, \apj, 708, 38

\bibitem[\protect\citeauthoryear{{Nagel}, {Espaillat}, {D'Alessio} \&
  {Calvet}}{{Nagel} et~al.}{2012}]{N2012}
{Nagel} E.,  {Espaillat} C.,  {D'Alessio} P.,    {Calvet} N.,  2012, \apj, 747,
  139

\bibitem[\protect\citeauthoryear{{Parks}, {Plavchan}, {White} \& {Gee}}{{Parks}
  et~al.}{2014}]{PPWG2014}
{Parks} J.~R.,  {Plavchan} P.,  {White} R.~J.,    {Gee} A.~H.,  2014, \apjs,
  211, 3

\bibitem[\protect\citeauthoryear{{Pichardo}, {Sparke} \& {Aguilar}}{{Pichardo}
  et~al.}{2005}]{PSA2005}
{Pichardo} B.,  {Sparke} L.,    {Aguilar} L.,  2005, \mnras, 359, 521

\bibitem[\protect\citeauthoryear{{Pichardo}, {Sparke} \& {Aguilar}}{{Pichardo}
  et~al.}{2008}]{PSA2008}
{Pichardo} B.,  {Sparke} L.,    {Aguilar} L.,  2008, \mnras, 391, 815

\bibitem[\protect\citeauthoryear{{Rattenbury}, {Wyrzykowski},
  {Kostrzewa-Rutkowska}, {Udalski}, {Koz{\l}owski}, {Szyma{\'n}ski},
  {Pietrzy{\'n}ski}, {Soszy{\'n}ski}, {Poleski}, {Ulaczyk} \& {et
  al.}}{{Rattenbury} et~al.}{2015}]{Rattenbury2015}
{Rattenbury} N.~J.,  {Wyrzykowski} {\L}.,  {Kostrzewa-Rutkowska} Z.,  {Udalski}
  A.,  {Koz{\l}owski} S.,  {Szyma{\'n}ski} M.~K.,  {Pietrzy{\'n}ski} G.,
  {Soszy{\'n}ski} I.,  {Poleski} R.,  {Ulaczyk} K.,    {et al.} 2015, \mnras,
  447, L31

\bibitem[\protect\citeauthoryear{{Shadmehri} \& {Khajenabi}}{{Shadmehri} \&
  {Khajenabi}}{2015}]{SK2015}
{Shadmehri} M.,  {Khajenabi} F.,  2015, \mnras, 447, 1439

\bibitem[\protect\citeauthoryear{{Siess}, {Dufour} \& {Forestini}}{{Siess}
  et~al.}{2000}]{S2000}
{Siess} L.,  {Dufour} E.,    {Forestini} M.,  2000, \aap, 358, 593

\bibitem[\protect\citeauthoryear{{Takakuwa}, {Saito}, {Lim} \&
  {Saigo}}{{Takakuwa} et~al.}{2013}]{Takakuwa2013}
{Takakuwa} S.,  {Saito} M.,  {Lim} J.,    {Saigo} K.,  2013, \apj, 776, 51

\bibitem[\protect\citeauthoryear{{Tal-Or}, {Faigler} \& {Mazeh}}{{Tal-Or}
  et~al.}{2015}]{TFM2015}
{Tal-Or} L.,  {Faigler} S.,    {Mazeh} T.,  2015, ArXiv e-prints

\bibitem[\protect\citeauthoryear{{van Boekel}, {Juh{\'a}sz}, {Henning},
  {K{\"o}hler}, {Ratzka}, {Herbst}, {Bouwman} \& {Kley}}{{van Boekel}
  et~al.}{2010}]{vJH2010}
{van Boekel} R.,  {Juh{\'a}sz} A.,  {Henning} T.,  {K{\"o}hler} R.,  {Ratzka}
  T.,  {Herbst} T.,  {Bouwman} J.,    {Kley} W.,  2010, \aap, 517, A16

\bibitem[\protect\citeauthoryear{{Williams} \& {Cieza}}{{Williams} \&
  {Cieza}}{2011}]{WC2011}
{Williams} J.~P.,  {Cieza} L.~A.,  2011, \araa, 49, 67

\bibitem[\protect\citeauthoryear{{Winn}, {Holman}, {Johnson}, {Stanek} \&
  {Garnavich}}{{Winn} et~al.}{2004}]{WHJ2004}
{Winn} J.~N.,  {Holman} M.~J.,  {Johnson} J.~A.,  {Stanek} K.~Z.,
  {Garnavich} P.~M.,  2004, \apjl, 603, L45

\bibitem[\protect\citeauthoryear{{Wolk}, {Rice} \& {Aspin}}{{Wolk}
  et~al.}{2013}]{WRA2013}
{Wolk} S.~J.,  {Rice} T.~S.,    {Aspin} C.,  2013, \apj, 773, 145

\bibitem[\protect\citeauthoryear{{Zucker}, {Mazeh} \& {Alexander}}{{Zucker}
  et~al.}{2007}]{ZMA2007}
{Zucker} S.,  {Mazeh} T.,    {Alexander} T.,  2007, \apj, 670, 1326

\end{thebibliography}

\end{document}